\documentclass[twocolumn,aps,prd,  
               preprintnumbers,numbers,sort&compress,
               nofootinbib, showpacs, colorlinks,
               linkcolor=blue, citecolor=blue]{revtex4-2} 

\usepackage{graphicx}
\usepackage{dcolumn}
\usepackage{bm}
\usepackage{hyperref}
\usepackage{xcolor}

\usepackage{amsmath}
\usepackage[caption=false]{subfig}
\usepackage{gensymb}

\def\ra{\rangle}
\def\la{\langle}
\def\rmd{\mathrm{d}}  

\newcommand{\be}{\begin{eqnarray}}
\newcommand{\ee}{\end{eqnarray}}
\newcommand{\beq}{\begin{equation}}
\newcommand{\eeq}{\end{equation}}
\newcommand{\exclude}[1]{}

\newcommand{\rev}[1]{#1} 

\exclude{

}
\graphicspath{{./Figures/}} 

\begin{document}
   
\title{Unidentified falling objects in the LHC as dark matter signals}
\author{Xunyu Liang}
\email{xunyul@phas.ubc.ca}
\author{Ariel Zhitnitsky}
\email{arz@phas.ubc.ca}
\affiliation{Department of Physics and Astronomy, University of British Columbia, Vancouver, V6T 1Z1, BC, Canada}
     
\begin{abstract}
     







Unidentified Falling Objects (UFOs) refer to 
sporadic
beam losses observed during LHC operation. The prevailing explanation is that micrometer-sized dust particles released from the beam screen produce beam losses through interactions with the protons. However, the release mechanism of these particles remains unknown. We propose that roughly $(1-10)\%$ of UFOs may be caused by axion quark nuggets (AQNs), macroscopic dark matter (DM) candidates with masses of order $(5-1000){\rm\,g}$. The AQN model naturally relates the dark- and visible-matter abundances ($\Omega_{\rm DM}\sim\Omega_{\rm visible}$) and provides a mechanism for generating the baryon–antibaryon asymmetry, with DM composed of both matter and antimatter AQNs. When passing underground within approximately $100{\rm\,km}$ of the LHC, an antimatter AQN generates acoustic waves strong enough to trigger multiple UFO events within 2\,s. If three correlated UFOs (placed at different locations along the LHC ring) are detected, the signal-to-noise ratio can exceed 5 across the entire allowed AQN mass range for a measurement time of about 360 hours. Practically, the LHC can serve as a large broadband acoustic detector for AQNs.


\end{abstract}

\maketitle





\section{Introduction}\label{sec:introduction}

Since 2010, it has been recognized that the so-called Unidentified Falling Objects (UFOs) may pose a significant limitation to LHC performance \cite{Baer:1379150,Baer:1493018}. A typical UFO appears as a sharp beam loss of up to $10^{8}$ protons occurring within a few milliseconds \cite{Lindstrom:2020hks}. UFOs are commonly attributed to micrometer-sized dust particles released from the beam screen that become attracted to the proton beams and produce beam losses via inelastic proton–nucleus collisions \cite{baer_2013_w6p20-zcn10}. However, the mechanism that releases these dust particles remains an open question \cite{Belanger:2020ufo,Lindstrom:2020hks}. UFO events are recorded by beam loss monitors (BLMs) and have been studied in detail, including analyses of multiplicity, time scales, and spatial distribution around the ring. The principal motivation for these studies has been to identify correlations between UFOs and beam parameters in order to improve LHC performance. Despite substantial mitigation efforts in recent years, dust-induced beam losses continue to affect LHC operation \cite{Lechner:2024olj}.

The motivation of the present work differs markedly from the operational goal discussed above of improving LHC performance. Instead, we propose that a small subset of UFO events could be initiated by dark matter (DM) in the form of axion quark nuggets (AQNs) \cite{Zhitnitsky:2002qa} passing underground within approximately $100{\rm\,km}$ of the LHC. The passage of a nearby antimatter AQN generates acoustic shock waves that can excite mechanical vibrations throughout the ring and potentially trigger UFO events. Identifying such DM-induced UFOs would represent a novel and unconventional method of DM detection.

We emphasize that the AQN model was not developed to explain UFOs; it was originally proposed to account for the observed similarity $\Omega_{\rm DM}\sim\Omega_{\rm visible}$ between dark and visible matter abundances \cite{Zhitnitsky:2002qa}. This similarity is a robust prediction of the framework and is largely insensitive to model parameters \cite{Ge:2017idw}. Moreover, the AQN scenario addresses the baryon-antibaryon asymmetry via asymmetric charge separation during nugget formation, resulting in a population that includes both matter and antimatter AQNs \cite{Liang:2016tqc,Ge:2017ttc,Ge:2019voa}. The model has also been invoked to explain a variety of puzzling observations spanning vastly different scales, from phenomena in the early Universe \cite{Flambaum:2018ohm} and at galactic scales \cite{Sekatchev:2025ixu} to solar \cite{Zhitnitsky:2017rop,Raza:2018gpb,Ge:2020xvf} and terrestrial effects \cite{Budker:2020mqk,Zhitnitsky:2020shd,Liang:2021wjx,Zhitnitsky:2022swb,Liang:2021rnv,Zhitnitsky:2024jnk}.

In fact, the idea that some UFO events could be related to AQNs was suggested in \cite{Zioutas:2024lgi}. However, that work did not propose a concrete mechanism for how an AQN could produce such events. The present work fills this gap by identifying a specific mechanism that can release the dust particles from the beam screen. These dust particles may then interact with the beams and be detected by the approximately 4000 BLMs distributed around the ring. The resulting AQN-induced UFO events have distinctive signatures and should be distinguishable from most other UFO events, as will be detailed in this work.

The title of this work includes two seemingly contradictory terms. ``UFO'' denotes events that can produce large, observable disturbances at the LHC (e.g., dumps or quenches), while ``DM'' ordinarily refers to invisible, effectively noninteracting matter on macroscopic scales. This contradiction is only apparent: there is a deep connection between the two. To clarify that connection, we briefly overview UFOs and the AQN model in Sec. \ref{sec:UFO events at the LHC} and Sec. \ref{AQN}, respectively. Starting in Sec. \ref{sect:proposal}, we explain how these two, apparently different entities (UFOs and AQNs) could nevertheless be closely related. We present our proposal, including the event rate, signal properties, and the LHC detection sensitivity, in Secs. \ref{sect:rate}, \ref{sect:detection}, and \ref{sec:Expected sensitivity}, respectively. Sec. \ref{sect:conclusion} summarizes our conclusions and discusses AQN-induced terrestrial phenomena similar to those described in this paper.



\section{UFO events at the LHC}
\label{sec:UFO events at the LHC}


The UFO signals refer to sporadic beam loss spikes that occur throughout the LHC. Most of these UFOs are characterized as regular UFOs, which have an asymmetric Gaussian profile of beam losses. Other types of UFOs, such as injection kicker magnets (MKI), unidentified lying objects (ULO), and 16L2, are observed less frequently \cite{Lindstrom:2020hks,Belanger:2020ufo}. The study of UFOs has continued since the onset of high-intensity beam operations at the LHC as these events can trigger beam dumps, magnetic quenches, and even cause damage to the LHC. UFOs are believed to be caused by micrometer-sized dust particles that enter the beam and interact with the protons \cite{baer_2013_w6p20-zcn10,Lechner:2021shq}. However, the precise release mechanism remains unknown. We will review the properties of UFOs and their release mechanism in subsection \ref{items} and \ref{subsec:UFO release mechanism}. 

\subsection{UFO properties}
 \label{items}

The properties of UFOs are discussed extensively in many documents, such as Refs. \cite{baer_2013_w6p20-zcn10,Belanger:2020ufo}. In this subsection, we present a brief overview that is relevant to this work.

The beam loss profile of regular UFOs appears as an asymmetric Gaussian-shaped spike over time, typically lasting between 0.1 and a few milliseconds \cite{Lindstrom:2020hks}. Regular UFOs serve as the primary experimental evidence supporting the UFO hypothesis \cite{Belanger:2020ufo}.

MKI UFOs have a beam loss profile very similar to that of regular UFOs. These events generally occurred within 30 minutes after the last injection \cite{baer_2013_w6p20-zcn10}. There is a clear correlation between the occurrence of MKI UFOs and pulsing MKIs. It is believed that $\rm Al_2O_3$ particles from the ceramic tube of the MKIs could be charged by electron clouds, and their release may occur due to vibrations or the electric field generated by the MKI pulse \cite{baer_2013_w6p20-zcn10}. 

A ULO is a contaminant, such as a strip of plastic originating from the beam pipe wrapping \cite{Mirarchi:2019mre}, found at the bottom of the beam pipe. This object’s interaction with the beam triggers UFO beam losses. The beam loss associated with ULOs is periodic, occurring approximately every 89$\,\mu$s. When the ULO is removed, the UFO beam loss ceases.

The 16L2 UFO represents a new type of beam loss. Unlike regular UFOs, persistent beam losses arise (historically, in cell 16L2) following the initial UFO-like beam-loss spike \cite{Lechner:2018fem}. The 16L2 UFO can be explained by a detached macro-particle of solid nitrogen/oxygen that interacts with the beam and undergoes phase transition from solid to gas \cite{Mether:2019xjr}.

Similar to the 16L2 UFO, UFO events can occur consecutively over time at different locations. Such a burst of UFOs is sometimes called a ``burst sequence'' or ``UFO storm''; see e.g. Refs. \cite{Myers:2013pzr,baglin2015other}. The main purpose of this work is to investigate the correlation of \textit{multiple independent} UFOs occurring closely together in time (within milliseconds to seconds) but at widely different locations. As we will demonstrate, this correlated burst of UFOs could serve as a significant signal for a macroscopic DM candidate, specifically the AQNs.





\subsection{UFO release mechanism}
\label{subsec:UFO release mechanism}
The UFO evolution in the LHC involves several stages: the formation, charging, and release of dust particles \cite{Belanger:2020ufo}. Typically, micrometer-sized dust particles can form and become charged on the beam screens within a few minutes during LHC operation. Given the several hours of daily operation, the formation of dust is quite likely. These dust particles typically acquire large negative surface charges ranging from $10^3$ to $10^4 e$ for particles sized between 1 and 10 micrometers.
 This characteristic is a well expected
feature of UFO evolution. What represents a mystery of UFOs is the release mechanism. Or to be formulated more precisely: what could trigger and initiate 
 the release of UFOs
from the beam screen? 
This work proposes a specific mechanism 
  that could serve as a trigger for the release of UFO particles.

\begin{figure}[h]               
  \centering
\includegraphics[width=1\linewidth]{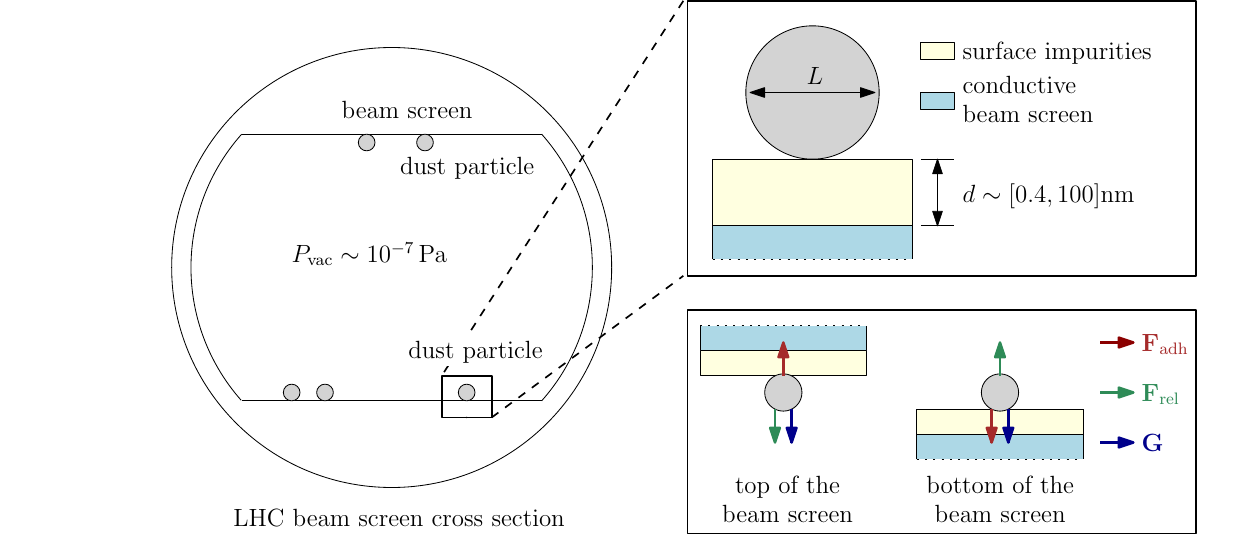}
  \caption{
  UFO release mechanism in the LHC. Micrometer-sized dust particles can form on the beam screens within several minutes during the LHC operation. Typically, these particles settle on the bottom of the beam screen, although it is also possible for them to attach to the top. A dust particle is influenced by several forces: an adhesive force $\mathbf{F}_{\rm adh}$ that keeps it attached to the beam screen, a gravitational force $\mathbf{G}$ that is determined by its size $L$, and an unknown releasing force $\mathbf{F}_{\rm rel}$. When the releasing force becomes strong enough, the dust particle will be dislodged and can trigger a UFO event. Specifically, for particles that attach to the top of the beam screen, they can self-release due to gravitational forces once they reach a sufficient size of approximately $L\gtrsim50{\rm\,\mu m}$ \cite{Belanger:2020ufo}.}
  \label{fig:force}
\end{figure}

As illustrated in Fig. \ref{fig:force}, most dust particles settle at the bottom of the beam screen; however, they may also attach to the top. The forces acting on a dust particle include an adhesive force $\mathbf{F}_{\rm adh}$ that keeps it attached to the beam screen at a distance $d$, a gravitational force $\mathbf{G}$ that depends on the dust particle's size $L$, and an unknown releasing force $\mathbf{F}_{\rm rel}$. The scalar equation for the release of dust is as follows:
\begin{equation}
\begin{aligned}
&F_{\rm rel}(L,d)
=F_{\rm adh}(L,d)\pm G(L)\,,  \\
&\rev{
F_{\rm adh}
\equiv F_{\rm coh} + F_{\rm vdw} +F_{\rm ic}\,.
}
\end{aligned}
\end{equation}
\rev{
Here, the sign of the gravitational force depends on whether the particle is located at the bottom or the top of the beam screen. The adhesive force is a combination of the cohesive force ($F_{\rm coh}$), the van der Waals force ($F_{\rm vdw}$), and the image charge force ($F_{\rm ic}$):
\begin{equation}
\begin{aligned}
F_{\rm coh}
=2\pi\gamma L\,,\quad
F_{\rm vdw}
=\frac{A_{\rm H}L}{12d^2}\,,\quad
F_{\rm ic}
=\frac{Q^2}{4\pi\varepsilon_0(L+2d)^2}\,,  \\
\end{aligned}
\end{equation}
where $\gamma\sim(0.02-2){\rm\,J\,m^{-2}}$ is the surface energy per unit area required to break the molecular bonds, $A_{\rm H}\sim(1-10)\times10^{-20}{\rm\,J}$ is the Hamaker constant of the interaction, $\varepsilon_0$ is the vacuum permittivity, and $Q$ is the equilibrium charge of the dust particle. Depending on the model, the charge may be approximated as $Q\propto R$ or $Q\propto R^2$.
}

Specifically, for particles attached to the top of the beam screen, they can self-release due to gravitational forces once they reach a sufficient size of approximately $L\gtrsim50{\rm\,\mu m}$. In principle, a UFO event can be triggered by a dust particle attached to the top of the beam screen when a change in mass, charge, or ambient electromagnetic fields occurs. However, particles adhering to the top of the beam screen are rare and unstable.

The majority of dust particles reside at the bottom of the beam screen. For a typical dust particle size of $L\approx10{\rm\,\mu m}$, the critical force required to release a dust particle is estimated, based on Ref. \cite{Belanger:2020ufo}, to be:
\begin{equation}
\label{eq:F_rel crit}
F_{\rm rel,crit}(L=10{\rm\,\mu m},d)
\sim10^{-8}{\rm\,N}\,,
\end{equation}
assuming \rev{$d \gtrsim 10{\rm\,nm}$} in a realistic scenario.\footnote{\rev{Specifically, the critical condition does not alter for $d=3{\rm\,nm}$ and $d=100{\rm\,nm}$ for a $Q\propto R^2$ model in Ref. \cite{Belanger:2020ufo}.}} Correspondingly, the critical activation energy and the dust vibration velocity can be estimated as follows:
\begin{subequations}
\label{eqs:E_crit etc.}
\begin{equation}
\label{eq:E_crit}
E_{\rm crit}
\approx F_{\rm rel,crit} L
\sim10^6{\rm\,eV}\,,
\end{equation}
\begin{equation}
v_{\rm crit}
\approx \sqrt{\frac{E_{\rm crit}}{2L^3\rho_{\rm dust}}}
\sim0.5{\rm\,m\,s^{-1}}\,,
\end{equation}
\end{subequations}
where $\rho_{\rm dust}$ is the mass density of a dust particle, which is typically on the order of a few grams per cubic centimetre. Such a vibration velocity exceeds any known external mechanical vibrations by several orders of magnitude. This includes vibrations caused by earthquakes ($\sim20{\rm\,\mu m\,s^{-1}}$) \cite{Charrondiere:2018ony} and the pulsing of MKI magnets ($\sim{\rm mm\,s^{-1}}$) \cite{ballester2011vibration}. In this paper, we propose that AQNs can meet the critical conditions \eqref{eq:F_rel crit} and \eqref{eqs:E_crit etc.}, triggering the release of dust at the bottom of the beam screen.


  \section{The AQN   dark matter  model}\label{AQN}

  




 
 We overview the AQNs as a DM matter candidate in subsection \ref{dm}. The basic characteristics of AQNs are presented in subsection \ref{basics}. Lastly, specific features of the AQNs relevant to the present work is presented in \ref{AQN-dense}.
 
 

 \subsection{AQNs as dark matter}\label{dm}


In this subsection, we provide a brief explanation of the term  \textit{dark matter}, which is also featured in the title of this work. From a cosmological perspective, there is a fundamental distinction between dark and ordinary matter, beyond the obvious difference between invisible and visible substances.
The key parameter which enters all the cosmological observations is the corresponding cross section $\sigma$ (describing the coupling of DM with standard model particles) to mass $M_{\rm DM}$ ratio, which must be sufficiently small to play the role of the DM, see e.g. recent review  \cite{Tulin:2017ara}:
\be
\label{sigma/m}
\frac{\sigma}{M_{\rm DM}}\ll  1\frac{\rm cm^2}{\rm g}.
\ee
Weakly Interacting Massive Particles (WIMPs) clearly meet the criteria (\ref{sigma/m}) to qualify as DM particles. This is mainly due to their tiny cross section $\sigma$ and a typical mass $M_{\rm WIMP}\in( 10^2-10^3) {\rm\,GeV}$. In contrast, ordinary visible matter obviously does not satisfy the criteria (\ref{sigma/m}).   

In this work, we consider a fundamentally different type of DM that takes the form of macroscopically large composite objects of nuclear density, similar to Witten's quark nuggets \cite{Witten:1984rs,Farhi:1984qu,DeRujula:1984axn}.  These objects are referred to as AQNs, as discussed in the original paper \cite{Zhitnitsky:2002qa} and a brief overview of AQN construction \cite{Zhitnitsky:2021iwg}. AQNs do not interact significantly in a dilute environment on cosmological scales, making them viable DM candidates. The necessary condition \eqref{sigma/m} for AQNs is met when the ratio ratio $\sigma /M_{\rm AQN}\leq 10^{-10} {\rm cm^2}{\rm g^{-1}}$. This is because the typical density of quark nuggets is 15 orders of magnitude greater than that of ordinary visible matter $\sim \rm g\, cm^{-3}$. However, these same objects interact very strongly with materials when they come into contact with the Earth or other planets and stars, meaning they are not truly \textit{dark} on local scales (such as around stars or planets). This new paradigm differs dramatically from conventional DM models such as WIMPs, which are assumed to be weakly interacting fundamental particles at all scales, regardless of the environment. 
  
The main distinguishing feature of the AQN model, which is crucial for the current work, is its ability to consist of both matter and antimatter during the QCD transition. This contrasts with the earlier construction by Witten \cite{Witten:1984rs,Farhi:1984qu,DeRujula:1984axn}. The process of charge segregation, outlined in a brief overview \cite{Zhitnitsky:2021iwg}, plays a significant role in this. During the QCD transition in the early Universe, the dynamics of the $\cal CP$-odd axion field cause the separation of quarks from antiquarks; see next subsection for more details. This separation of baryon charges leads to the formation of quark nuggets and antiquark nuggets at similar, though not identical, rates.

 \subsection{Characteristics of AQNs}
 \label{basics}
 
 The original motivation for the AQN model can be explained as follows: for a brief overview of the AQN construction, refer to Ref. \cite{Zhitnitsky:2021iwg} for details. It is commonly assumed that the Universe began in a symmetric state with zero global baryon charge and later evolved into a state with a net positive baryon number through some baryon-number-violating processes, nonequilibrium dynamics, and $\cal{CP}$-violation effects, effectively adhering to the three well known Sakharov criteria.

The AQN proposal is an alternative to this commonly accepted scenario, when ``baryogenesis'' is replaced by an asymmetric charge separation  
in which the global baryon number of the Universe remains 
zero at all times.  This is the key element of the AQN construction.

In the AQN scenario, the DM density $\Omega_{\rm DM}$ is composed of both matter and antimatter AQNs, and will automatically (regardless of the axion mass $m_a$ or the misalignment angle $\theta_0$) take on the same order of magnitude as the visible density, $\Omega_{\rm visible}$. Both densities are proportional to one and the same fundamental dimensional parameter of the theory, $\Lambda_{\rm QCD}$. Therefore, the AQN model inherently resolves two fundamental problems in cosmology: it explains the baryon asymmetry of the Universe and accounts for the presence of DM with the proper density relation $\Omega_{\rm DM} \sim \Omega_{\rm visible}$, without the need to fit any parameters of the model.


Given the predominance of ordinary matter in the Universe, observable interactions will be dominated by the antimatter AQNs in most scenarios. For simplicity, we will hereafter refer to ``antimatter AQNs'' simply as ``AQNs''.


The primary fundamental parameter of this AQN framework is the average baryon charge (or equivalently, the mass) of nuggets. The strongest constraint on the mass of an AQN from direct nondetection is set by the IceCube Observatory, as detailed in Appendix A of Ref. \cite{Lawson:2019cvy}:  
\begin{equation}
\label{direct}
\begin{aligned}
\la M_{\rm AQN}\ra\approx m_p\la B\ra
\gtrsim5{\rm\,g}\quad\textrm{(direct nondetection)}
\,,
\end{aligned}
\end{equation}
where $\langle M_{\rm AQN}\rangle$ is the average mass of the AQNs, $\langle B\rangle$ is the average baryon charge of the AQNs, and $m_p$ is the mass of a proton. In terms of the baryon charge, the constraint \eqref{direct} is equivalent to $\langle B\rangle\gtrsim3\times10^{24}$. 

AQNs are significantly heavier objects, with masses exceeding $3\times10^{24}$\,GeV, compared to conventional WIMPs, which have masses on the order of $100\,$GeV. Consequently, the event rate for AQNs is dramatically lower than the typical event rate for conventional WIMPs at the same DM density $\rho_{\rm DM}$. 
The AQN flux can be estimated as follows \cite{Lawson:2019cvy}: 
\begin{equation}
\label{Phi1}
\begin{aligned}
\frac{  \rmd \Phi}{  \rmd A}
\approx\frac{\Phi}{4\pi R_\oplus^2}  
&\approx6.93\times10^{-3}{\rm\,km^{-2}\,yr^{-1}}
\left(\frac{100{\rm\,g}}{\langle M_{\rm AQN}\rangle}\right)  \\
&\rev{\quad\times\left(\frac{\rho_{\rm DM}}{0.3{\rm\,GeV\,cm^{-3}}}\right)
\left(\frac{v_{\rm AQN}}{220{\rm\,km\,s^{-1}}}\right)
}
\,,
\end{aligned}
\end{equation}
where $R_\oplus=6371\,$km is the radius of the Earth, the average 
AQN mass $\langle M_{\rm AQN}\rangle\approx100{\rm\,g}$ is the benchmark value being extracted from recent observation \cite{Sekatchev:2025ixu} (see discussion below for details),   and  $\Phi$ is the total  flux  of the AQNs   expressed in terms of the DM density 
$\rho_{\rm DM}$ on Earth \cite{Lawson:2019cvy}: 
\begin{equation}
\label{Phi}
\begin{aligned}
\Phi
&\approx3.54\times10^6{\rm yr^{-1}}
\left(\frac{100{\rm\,g}}{\langle M_{\rm AQN}\rangle}\right)
\left(\frac{\rho_{\rm DM}}{0.3{\rm\,GeV\,cm^{-3}}}\right) \\
&\rev{
\quad\times\left(\frac{v_{\rm AQN}}{220{\rm\,km\,s^{-1}}}\right)
}
\,,
\end{aligned} 
\end{equation}
Although not for the current proposal, for our future investigation of the AQN-induced UFO event rate, we need to understand the mass distribution of the AQNs. This distribution, denoted as $f(M_{\rm AQN})$, is defined as follows: 

Let $\rmd N/\rmd M_{\rm AQN}$ be the number of AQNs with mass in the range [$M_{\rm AQN}$, $M_{\rm AQN}+\rmd M_{\rm AQN}$]. The mean value of the AQN mass $\la M_{\rm AQN}\ra$ is given by  
\begin{equation}
\label{eq:f(B)}
\begin{aligned}
&\langle M_{\rm AQN}\rangle
=\int\rmd M_{\rm AQN}\, M_{\rm AQN}f(M_{\rm AQN})\,, \\
&f(M_{\rm AQN})\propto M_{\rm AQN}^{-\alpha}\,,
\end{aligned}
\end{equation}
where $f(M_{\rm AQN})$ is a  properly normalized distribution,  and $\alpha\approx (2-2.5)$  is the power-law index \cite{Raza:2018gpb}. The algebraic scaling (\ref{eq:f(B)}) is a generic characteristic of the AQN formation mechanism based on percolation theory  \cite{Ge:2019voa}. However, the power-law index  $\alpha$ and the specific mass $M_{\rm AQN}$ cannot be theoretically computed within strongly coupled QCD; instead, these parameters should be extracted from observations when a signal is detected and distinguished from noise. The corresponding value for $\langle M_{\rm AQN}\rangle$ must also satisfy the constraint (\ref{direct}).





As previously mentioned in this subsection, AQNs are macroscopic DM that interact with the Earth only rarely. Recent investigations into the electromagnetic signatures associated with AQNs on an astrophysical scale have provided valuable insights. Studies focusing on the early Universe, large scale structures, and the electromagnetic emissions from galaxies, such that the average mass of AQN is approximately $\langle M_{\rm AQN}\rangle\approx100{\rm\,g}$ and no more than $1{\rm\,kg}$ \cite{Majidi:2024mty,Sommer:2024iqp,Sekatchev:2025ixu,Majidi:2025ylh}. Assuming the local properties of AQNs are consistent with these global observations, we will adopt $\langle M_{\rm AQN}\rangle\approx100{\rm\,g}$ as our benchmark for all estimates in this work. This choice also serves as a nontrivial self-consistency check when we use this benchmark to estimate the characteristics of the AQN-induced UFOs in the following sections.

 \subsection{AQN impacts on Earth}\label{AQN-dense}


 
The computations regarding the interaction of AQN-visible matter were initially conducted in Ref. \cite{Forbes:2008uf}, focusing on the galactic environment, where the typical density of surrounding baryons is approximately $ n_{\text{galaxy}} \sim \, \text{cm}^{-3} $ and the internal temperature is in the range of $ T \sim \text{eV} $. To avoid repetition of certain formulas, we will reference recent papers \cite{Budker:2020mqk, Zhitnitsky:2024jnk, Zhitnitsky:2025bvy}, which provide a brief overview of the fundamental concepts related to the interaction when AQN reaches the Earth, the focus of this paper.
  
Currently, we would like to highlight some estimates regarding the features of AQN as they propagate through the Earth's atmosphere. In this context, the internal temperature of the AQN begins to increase, reaching approximately 20 keV. Consequently, the typical internal temperature of the nuggets in the Earth's atmosphere is estimated as follows \cite{Budker:2020mqk}:
 \be
 \label{T}
 T\approx 20{\rm\, keV}  \left(\frac{n_{\rm air}}{10^{21} ~{\rm cm^{-3}}}\right)^{\frac{4}{17}}\,,
 \ee 
where  typical density of surrounding baryons is   $n_{\rm air}\simeq 30N_m\simeq 10^{21}{\rm\,cm^{-3}}$, where $N_m\simeq 2.7\times10^{19}{\rm\,cm^{-3}}$ is the molecular density in the atmosphere, with each molecule containing approximately 30 baryons. 

Similar arguments also apply to an AQN propagating deep beneath the Earth's surface. In this case, the density is approximately $n_{\rm rock}\approx 10^{24}{\rm\,cm^{-3}}$. Consequently, the internal temperature of the AQN reaches much higher values, in the 
$T\approx 10^2{\rm\,keV}$ range, as per Eq.~(\ref{T}). This high internal temperature $T$ is crucial for estimating the basic characteristics of the shock waves generated by the supersonic motion of the AQN; see Sec.~\ref{subsec:Acoustic shock waves emitted by AQNs}. 
These shock waves will mechanically vibrate the LHC machine, consequently releasing dust particles from the surface and ultimately resulting in unique UFO events, as discussed in Sec.~\ref{sect:proposal}.







In short, the system contains antimatter nuggets that behave as DM objects in a dilute cosmological environment. When these nuggets enter the Earth's atmosphere, they undergo annihilation, producing extremely energetic events. In particular, these annihilation events may generate, among many other effects mentioned in concluding Sec. \ref{sect:conclusion}, very rare but significant phenomena such as shock waves that propagate long distances through the sky or deep underground. This phenomenon, observed for centuries worldwide without a clear physical explanation, is commonly known as \textit{skyquakes}. The study \cite{Budker:2020mqk} have proposed that skyquakes could result from antimatter AQNs traversing the atmosphere. 
\rev{
We briefly review this phenomenon in Appendix~\ref{sect:skyquakes} and related terrestrial observations in Appendix~\ref{app:Related terrestrial observations}.
}

\rev{

\subsection{Acoustic shock waves emitted by AQNs}
\label{subsec:Acoustic shock waves emitted by AQNs}

This subsection briefly reviews the work \cite{Budker:2020mqk} that investigates the acoustic shock waves emitted by an AQN in the atmosphere and underground. 

An AQN propagating in the atmosphere is similar to the case of meteor. In both scenarios, the motion of the object is supersonic and produce acoustic shock waves. We are specifically interested in the acoustic shock waves far away from the object. In the far-field limit\footnote{In the case of AQNs, the far-field limit corresponds to approximately $r\gg(0.3-10)$\,m; see the original paper \cite{Budker:2020mqk} for details. In our present scenario, where $r\sim100\,$km, this approximation is clearly satisfied.}, the acoustic shock wave emitted by a meteor has a well defined scaling relation between the distance and the pressure and frequency \cite{revelle1976meteor,silber2014optical,silber2015optical}:
\begin{equation}
\label{eqs:P r etc.}
P(r)
\propto r^{-\frac{3}{4}}\,,\qquad
\nu(r)
\propto r^{-\frac{1}{4}}\,,
\end{equation}
where $r$ is the distance from the meteor, $P$ and $\nu$ is the pressure and frequency of the acoustic shock wave. The acoustic wave is assumed to follow a similar relation defined above. The scaling relation fails in a near-field distance $L_{\rm abs}$, which is approximately the photon absorption distance of the surrounding environment. In Ref. \cite{Budker:2020mqk}, this is estimated to be:
\begin{equation}
L_{\rm abs}
\approx\left\{
\begin{aligned}
&5{\rm\,m}\,, &\textrm{in the air}; \\
&\eta\cdot2{\rm\,cm}\,, &\textrm{in the rock}; \\
\end{aligned}
\right.
\end{equation}
where $\eta$ is the absorption parameter such that $\eta\approx1$ in pure silicon for an AQN temperature $T\sim100\,$keV underground; see the corresponding discussion under Eq. \eqref{T}.

Within the region defined by $L_{\rm abs}$, the system is approximately in thermal equilibrium with a constant pressure, rather than following the scaling relation \eqref{eqs:P r etc.}. In the atmosphere, the thermal energy $U$ of the diatomic ideal gas is given as:
\begin{equation}
\rmd U=\frac{2}{5}P\rmd V\,,\qquad
\rmd V=\pi L_{\rm abs}^2\rmd l\,,
\end{equation}
where $\rmd V$ is the volume of an infinitesimal cylinder along the direction of the propagating AQN $\rmd l$, within the thermal equilibrium distance $L_{\rm abs}$. 

In thermal equilibrium, the internal energy is equal to the annihilation energy $E_{\rm ann}$ (between the antimatter AQN and its surrounding medium) released by the AQN:
\begin{equation}
\rmd U=\rmd E_{\rm ann}
\end{equation}
\begin{equation}
\frac{\rmd E_{\rm ann}}{\rmd l}
=0.1\xi\times2m_pc^2\sigma n_{\rm env}\,,
\end{equation}
where the prefactors $0.1$ and $\xi$ characterize the efficiency of annihilation and photon emission, respectively; $m_p$ is the proton mass; $c$ is the speed of light in vacuum; $\sigma\approx5.24\times10^{-13}{\rm\,m}^2\left(\frac{m_{\rm AQN}}{100{\rm\,g}}\right)^{2/3}$ is the geometrical cross section of the AQN; $n_{\rm env}$ is the baryon number density of the surrounding environment of the AQN. In this expression, we assume that each $p\bar{p}$ pair annihilation releases an energy of $2m_pc^2$. The emission efficiency is suppressed in a high-density environment (e.g. underground) due to drastic decrease of positron density on the AQN surface, such that $\xi$ is estimated to be:
\begin{equation}
\xi
\sim\left\{
\begin{aligned}
&1\,, &\textrm{in the air}; \\
&10^{-2}\,, &\textrm{in the rock}. \\
\end{aligned}
\right.
\end{equation}
As discussed in Ref.~\cite{Budker:2020mqk}, the acoustic wave propagating in the rock may have an additional exponential attenuation $e^{-X(r)}$ as a function of $r$ due to sound absorption. This is equivalent to the transformation $\xi\rightarrow\xi e^{-X(r)}$. However, the physics is complicated in solids and difficult to evaluate. At least, it is certain that such sound absorption is weak in liquids for $r\lesssim100\,$km. We 
will assume the sound absorption in kHz frequency bands is weak in solids in this work, as in the original work~\cite{Budker:2020mqk}.

The equations above imply that the thermal pressure at $r=L_{\rm abs}$ is
\begin{subequations}
\begin{equation}
P(L_{\rm abs})
\approx\frac{2}{5\pi L_{\rm abs}^2}\frac{\rmd E_{\rm ann}}{\rmd l}\,,
\end{equation}
\begin{equation}
\nu(L_{\rm abs})
\approx\frac{c_{\rm s}}{L_{\rm abs}}\,,
\end{equation}
\end{subequations}
where the second equation $\nu(L_{\rm abs})$ is a trivial relation from dimensional analysis. Comparing with the scaling relation \eqref{eqs:P r etc.}, we conclude
\begin{subequations}
\begin{equation}
P(r)
\approx P(L_{\rm abs})
\left(\frac{L_{\rm abs}}{r}\right)^{\frac{3}{4}}\,,
\end{equation}
\begin{equation}
\nu(r)
\approx\nu(L_{\rm abs})
\left(\frac{L_{\rm abs}}{r}\right)^{\frac{1}{4}}\,.
\end{equation}
\end{subequations}
Based on the parameters chosen above, we estimate the pressure and frequency in the air:
\begin{subequations}
\label{eqs:P_air etc.}
\begin{equation}
\begin{aligned}
P_{\rm air}(r)
&\sim0.05{\rm\,Pa}
\left(\frac{100{\rm\,km}}{r}\right)^{\frac{3}{4}}
\left(\frac{\langle M_{\rm AQN}\rangle}{100{\rm\,g}}\right)^{\frac{2}{3}} \\
&\quad\times\left(\frac{n_{\rm air}}{10^{21}{\rm\,cm^{-3}}}\right)
\,,
\end{aligned}
\end{equation}
\begin{equation}
\nu_{\rm air}(r)
\sim6{\rm\,Hz}
\left(\frac{100{\rm\,km}}{r}\right)^{\frac{1}{4}}
\left(\frac{c_{\rm s}}{340{\rm\,m\,s^{-1}}}\right)
\left(\frac{1}{\eta}\right)^{\frac{3}{4}}\,.
\end{equation}
\end{subequations}
and in the rock:
\begin{subequations}
\label{eqs:P_rock etc.}
\begin{equation}
\begin{aligned}
P_{\rm rock}(r)
&\sim5\times10^2{\rm\,Pa}
\left(\frac{100{\rm\,km}}{r}\right)^{\frac{3}{4}}
\left(\frac{\langle M_{\rm AQN}\rangle}{100{\rm\,g}}\right)^{\frac{2}{3}} \\
&\quad\times
\left(\frac{n_{\rm rock}}{10^{24}{\rm\,cm^{-3}}}\right)
\left(\frac{\xi}{10^{-2}}\right)
\left(\frac{1}{\eta}\right)^{\frac{5}{4}}\,,
\end{aligned}
\end{equation}
\begin{equation}
\nu_{\rm rock}(r)
\sim4{\rm\,kHz}
\left(\frac{100{\rm\,km}}{r}\right)^{\frac{1}{4}}
\left(\frac{c_{\rm s}}{4{\rm\,km\,s^{-1}}}\right)
\left(\frac{1}{\eta}\right)^{\frac{3}{4}}\,.
\end{equation}
\end{subequations}


One can verify that the numerical values and their dependence on variables are consistent with the original paper~\cite{Budker:2020mqk}, albeit with minor differences due to small differences in normalization factors such as the AQN benchmark mass $\langle M_{\rm AQN}\rangle$.  As previously mentioned, the evaluation of $P(r)$ and $\nu$ provides only an order-of-magnitude estimate. The values of $P(r)$ and $\nu$ may vary based on the choice of parameters, such as $n_{\rm env}$, $c_{\rm s}$, $\eta$, and $\xi$. In the following analysis, we will choose slightly different numerical values than those in Eqs.~\eqref{eqs:P_air etc.} and \eqref{eqs:P_rock etc.} to match those from the original paper~\cite{Budker:2020mqk} for the sake of consistency.


}

\section{Proposal: LHC as the DM detector}
\label{sect:proposal}



The basic idea of our proposal has already mentioned in the earlier sections. We are now ready to formulate our proposal explicitly. The study  \cite{Budker:2020mqk} suggests that typical AQN-induced events, which we identify as skyquakes (as reviewed in Appendix \ref{sect:skyquakes}) with  $M_{\rm AQN}\approx (10-100)\,$g, are not intense enough to be detected without dedicated scientific instruments similar to  ELFO. In other words, the majority of AQN events will go unnoticed at large distances (around $10^2$ km)  due to relatively modest emitted energy in acoustic frequency bands. 

However,  these same typical AQN  events with  $M_{\rm AQN}\approx (10-100)\,$g are still sufficiently intense to cause mechanical vibrations in the LHC ring, which can release the dust particles from its surface.  Once these dust particles are released, they get attracted by the proton beams and can trigger the UFO events, 
as reviewed in Sec. \ref{sec:UFO events at the LHC}. 
We have new information to add regarding this latter stage in the evolution of the dust particles.  Our contribution is relatively modest, focusing only on the mechanism of releasing dust particles from the surface, which represents the very first stage of the UFO events.  

Nonetheless, these events are unique in many ways and can be easily distinguished from the majority of conventional UFO events, as argued in Sec. \ref{sect:detection}. It is important to note that these events are triggered by DM objects, making the LHC an ideal detector to study the macroscopic DM in the form of AQN-induced UFO events.

It is clear that only a small subset of the UFO events could be attributed to AQN-induced events. This subset likely accounts for only  $(1-10)\%$ of all UFO events, as estimated in Sec. \ref{sect:rate}, and is related to the vibrations caused by nearby passing AQNs.

\begin{figure}[h]               
  \centering
\includegraphics[width=1\linewidth]{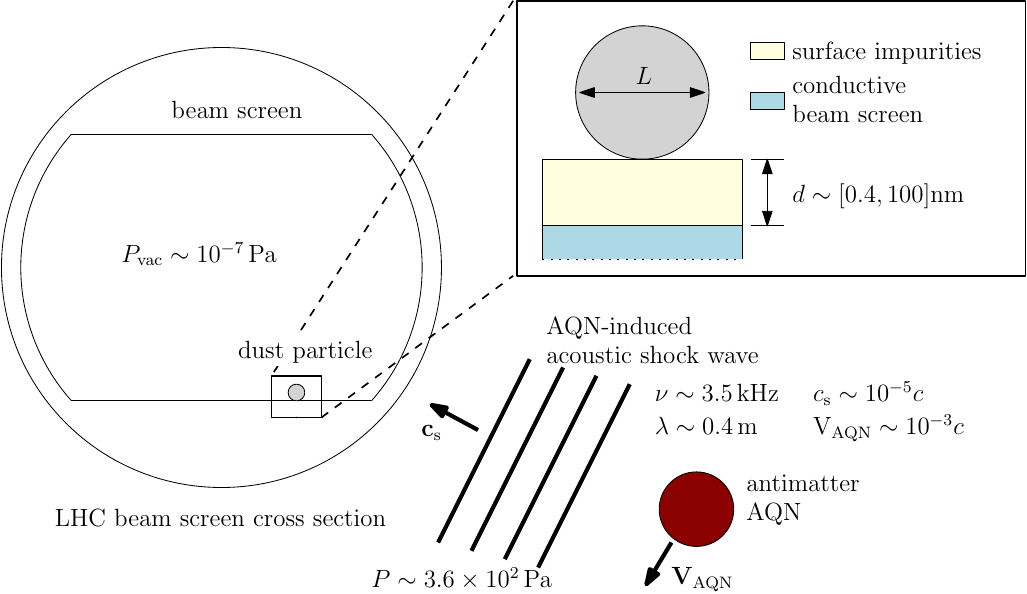}
  \caption{A dust particle located at the bottom of the beam screen receives an instantaneous mechanical impulse from an external acoustic shock wave induced by an (antimatter) AQN. An underground-propagating AQN, within approximately $100\,$km of the LHC tunnel, generates an overpressure of a few hundred Pascals and a frequency of a few kHz. The dust particle as a whole experiences the force coherently. }
  \label{fig:dust}
\end{figure}

To demonstrate that the passing AQNs can cause sufficiently strong vibration of the LHC machine, we need to estimate the overpressure $P(r)$ and frequency $\nu(r)$ of the shock wave at distance $r$ from the source; see Fig. \ref{fig:dust}. We start with an estimate for the overpressure at distance $r$ from the AQN for acoustic waves propagating underground, where the LHC ring is located:
\begin{subequations}
\label{eq:overpressure}
\begin{equation}
P(r)
\sim 3.6\times10^2{\rm\,Pa}
\left(\frac{100{\rm\,km}}{r}\right)^{\frac{3}{4}} 
\left(\frac{\langle M_{\rm AQN}\rangle}{100{\rm\,g}}\right)^{\frac{2}{3}}\,,
\end{equation}
\begin{equation}
\nu(r)
\sim3.5{\rm\,kHz}\left(\frac{1}{\eta}\right)^{\frac{3}{4}}
\left(\frac{100{\rm\,km}}{r}\right)^{\frac{1}{4}}\,.
\end{equation}
\end{subequations}
\rev{
Here we choose the numerical values to match those from the original 
paper~\cite{Budker:2020mqk}, rather than those in Eqs.~\eqref{eqs:P_rock etc.}, for the sake of consistency as noted in Sec.~\ref{subsec:Acoustic shock waves emitted by AQNs}.
}

The estimate (\ref{eq:overpressure}) indicates that the acoustic shock wave generates an additional force on a dust particle of area $A=L^2$ at distance $r$: 
\be
\label{eq:force}
\begin{aligned}
\begin{aligned}
F(r)
\approx P(r)\,A
&\approx3.6\times10^{-8}{\rm\,N}
\left(\frac{100{\rm\,km}}{r}\right)^{\frac{3}{4}} \\
&\quad\times
\left(\frac{\langle M_{\rm AQN}\rangle}{100{\rm\,g}}\right)^{\frac{2}{3}}
\left(\frac{L}{10{\rm\,\mu m}}\right)^2\,,
\end{aligned}
\end{aligned}
\ee
which satisfies the critical condition \eqref{eq:F_rel crit}. In the estimate (\ref{eq:force}), it is assumed that the dust particles are attached to a solid surface of the beam screen where $P(r)$ applies, while opposite side of the dust particle faces the vacuum with $P_{\rm vac}\approx 10^{-7}{\rm\,Pa}$. Another assumption made in the estimate (\ref{eq:force}) is that the wavelength of the acoustic wave, $\lambda(r)=c_s/\nu(r)$, is sufficiently large such that the dust particle $L$ as a whole experiences the force with the same sign (in phase). This assumption holds true since $L\ll \lambda$. 

Furthermore, the force $F(r)$ gives an impulse on the dust particle particle. The change of momentum for the dust particle is given by:
\begin{equation}
\label{eq:Delta p}
\Delta p
\approx\int\rmd t\,F(r)
\approx\frac{1}{4\nu(r)} F(r)\,,
\end{equation}
where $\Delta p$ is estimated as a root mean square average of the momentum change, and we assume the shock wave is sinusoidal and only the first half of the oscillation cycle contributes to the impulse.
Taking $M\approx\rho L^3$ and $\rho_{\rm dust}\approx3{\rm\,g\,cm^{-3}}$, we can estimate the change in kinetic energy $\Delta E_{\rm k}$:
\begin{equation}
\label{eq:Delta E_k}
\begin{aligned}
\Delta E_{\rm k}
=\frac{\Delta p^2}{2M}
&=6.70\times10^6{\rm\,eV}
\left(\frac{100{\rm\,km}}{r}\right)
\left(\frac{3{\rm\,g\,cm^{-3}}}{\rho_{\rm dust}}\right)  \\
&\quad\times
\left(\frac{\langle M_{\rm AQN}\rangle}{100{\rm\,g}}\right)^{\frac{4}{3}}
\left(\frac{L}{10{\rm\,\mu m}}\right)\,,  \\
\end{aligned}
\end{equation}
and the corresponding vibration velocity is:
\begin{equation}
\label{eq:Delta v}
\begin{aligned}
\Delta v
&=0.846{\rm\,m\,s^{-1}}
\left(\frac{100{\rm\,km}}{r}\right)^{\frac{1}{2}} 
\left(\frac{3{\rm\,g\,cm^{-3}}}{\rho_{\rm dust}}\right)  \\
&\quad\times
\left(\frac{\langle M_{\rm AQN}\rangle}{100{\rm\,g}}\right)^{\frac{2}{3}}
\left(\frac{10{\rm\,\mu m}}{L}\right)\,.  \\
\end{aligned}
\end{equation}
One can check $\Delta E_{\rm k}$ and $\Delta v$ both satisfy the critical release conditions of UFOs [see Eqs. \eqref{eqs:E_crit etc.}] as expected. 

In short conclusion, the overpressure \eqref{eq:overpressure} from passing AQN at distance $r\sim100{\rm\,km}$ provides more than sufficient mechanical vibration to release the dust particles from the surface.

The dust particles released will interact with the beam to generate UFO events, as discussed in Sec. \ref{sec:UFO events at the LHC}. This forms the basis of our proposal for releasing dust particles through the AQN-induced mechanism. 







It is essential to emphasize that the AQN-induced mechanism is significantly different from conventional UFO mechanisms. In conventional mechanisms, the vibration source and the released dust are localized. In contrast, the vibrations caused by an AQN can lead to release of dust particles throughout the entire machine ring, including areas such as arcs and collimation regions. The correlation time for these multiple UFOs, or a UFO burst, can range from milliseconds to several seconds. This variation depends on the trajectory of the AQN and the separation distance between two correlated UFOs, as will be discussed in Section \ref{sect:detection}.

We consider the energy scale determined by Eq. \eqref{eq:Delta E_k} is relatively low when rescaled to the energy per atom. Specifically, assuming a typical atomic number $A$ for a dust particle around $A\sim 10^{16}$, we conclude that the energy increment per atom from the acoustic wave is approximately $10^{-9}\,$eV/atom, which is an extremely low energy transfer. 

Nevertheless, this energy transfer can still be measured and recorded, thanks to the LHC's design, which features very low pressure in the beam region and 4000 BLMs installed along the ring. This makes the LHC unique, as it allows for the study of correlations between different UFOs in various locations, triggered and initiated by the same AQN passing 100 km away from the LHC. One could think of the LHC as a large acoustic DM detector due to its considerable size, especially when compared to typical DM detectors, and the large number of BLMs along the LHC ring.

\rev{

\section{Transfer function}\label{sect:transfer}

In the previous estimate, we assumed that the dust particle experiences the AQN-induced pressure wave as if it were applied directly to the beam screen in the LHC tunnel. However, the beam screen is linked through a complex system including the cold mass, magnets, support posts, thermal shield, vacuum vessel, and floor jacks \cite{Bourcey:2004un}. Such a mechanical system typically exhibits resonance frequencies in the range of tens of Hz \cite{Guinchard:2018lxq}. At first glance, it seems unlikely that a 3.5\,kHz acoustic wave could transmit through the tunnel wall to a dust particle on the LHC beam screen without a significant loss in pressure amplitude.\footnote{\rev{ We gratefully thank the anonymous referee for requesting to elaborate on this matter in the present work. We hope this section removes possible confusion between common terminology on acoustic signals in Hz versus kHz frequency bands. 
}}

However, the transmission mechanism differs fundamentally between low-frequency ($\nu \ll \nu_{\rm c}$) and high-frequency ($\nu \gg \nu_{\rm c}$) vibrations, where the characteristic frequency $\nu_{\rm c}$ is defined as:
\begin{equation}
\label{eq:nu_c}
\nu_{\rm c} 
\approx \frac{c_{\rm s}}{15{\rm\,m}}
= 0.27{\rm\,kHz} \,,
\end{equation}
and 15\,m is the characteristic length of an LHC superconducting magnet \cite{FernandezCano:2006rn}. At low frequencies, the acoustic wave acts on the entire system coherently, resulting in bulk displacement. Conversely, at high frequencies, bulk motion is absent and the acoustic wave propagates via internal stress.
The fact that kHz frequency acoustic signals can propagate hundreds of km in rock without significant absorption is a well-known phenomenon. In the context of the present work, the mysterious ELFO event (which was identified in Ref. \cite{Budker:2020mqk} with a very energetic AQN event) was accompanied by audible seismic signals in kHz frequency bands. It had been recorded by a network of seismic stations located hundreds of km apart and capable of measuring audible signals in kHz frequency bands.

\begin{figure*}[t]
    \centering
    \includegraphics[width=0.9\textwidth]{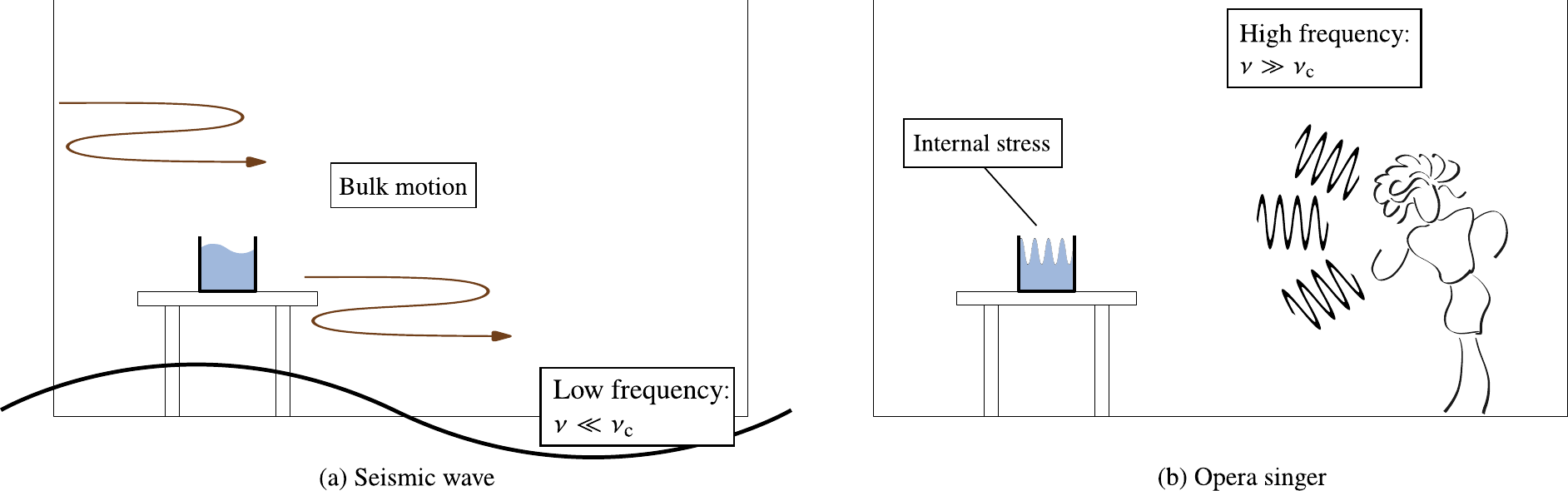}
    \caption{\rev{An analog of low- vs high-frequency response. Left panel (a): In the low-frequency scenario, the seismic wave moves a house. Everything (table, glass, and water) moves together coherently and slowly. Right panel (b): In the high-frequency scenario, an opera singer hits a high note. Nothing moves this time. But the glass experiences internal stress due to high-frequency pressure waves.}}
    \label{fig:analog}
\end{figure*}

In the text below we explain the differences between low-frequency and high-frequency modes in an intuitive way using the analogy illustrated in Fig.~\ref{fig:analog}. In the low-frequency scenario (a), everything moves coherently and no internal stress builds up. In the high-frequency scenario (b), the glass experiences internal stress without any macroscopic motion --- analogous to how a dust particle on the beam screen can be dislodged without the tunnel wall experiencing any significant displacement. In the case of the LHC, a tiny activation force of approximately $10^{-8}{\rm\,N}$ would be sufficient to trigger a UFO event, while the total force applied to the external tunnel wall by the pressure wave is about $360{\rm\,Pa}\cdot\pi\cdot 3.8{\rm\,m}\cdot 15{\rm\,m}=6.4\times10^4{\rm\,N}$. Even if $10^{-12}$ of the total energy survives and is transmitted to the dust particle, the dust particle may be dislodged from the beam screen.

The remaining question is whether the pressure amplitude is preserved and transmitted through the complex LHC system. To estimate the pressure transfer function\footnote{\rev{As discussed earlier in this subsection, we should  estimate the \textit{pressure} transfer function, rather than the \textit{displacement} transfer function, for high-frequency waves. The displacement transfer function is only 
applicable to low-frequency waves \cite{Schaumann:2023gst,Guinchard:2018lxq,Verdier:2000df}.}}, we model the transmission through the following layers in order: the concrete tunnel wall, floor jacks, vacuum vessel, support posts, cold mass, and finally the beam screen; see Fig.~\ref{fig:tunnel}. The corresponding properties and dimensions are summarized in Table~\ref{tab:components}.

When an acoustic wave propagates from one medium $i$ to a different medium $j$, 
the pressure transmission coefficient is given by:
\begin{equation}
\label{eq:T_ij}
\mathcal{T}_{ij}
=\frac{2Z_j}{Z_i+Z_j}\,,\qquad
Z_i
=\rho_ic_{{\rm s},i}\,,
\end{equation}
where $Z_i$ is the acoustic impedance, $\rho_i$ is the mass density of the medium, and $c_{{\rm s},i}$ is the sound speed in the medium; see e.g.\ textbooks 
\cite{landau-fluid,Benenson:2002qu}, and Appendix~\ref{app:Pressure transmission coefficient}. Here and in what follows, the subscript indices $i,j$ follow the labels in Table~\ref{tab:components}.

\begin{figure}[h]
    \centering
    \includegraphics[width=0.48\textwidth]{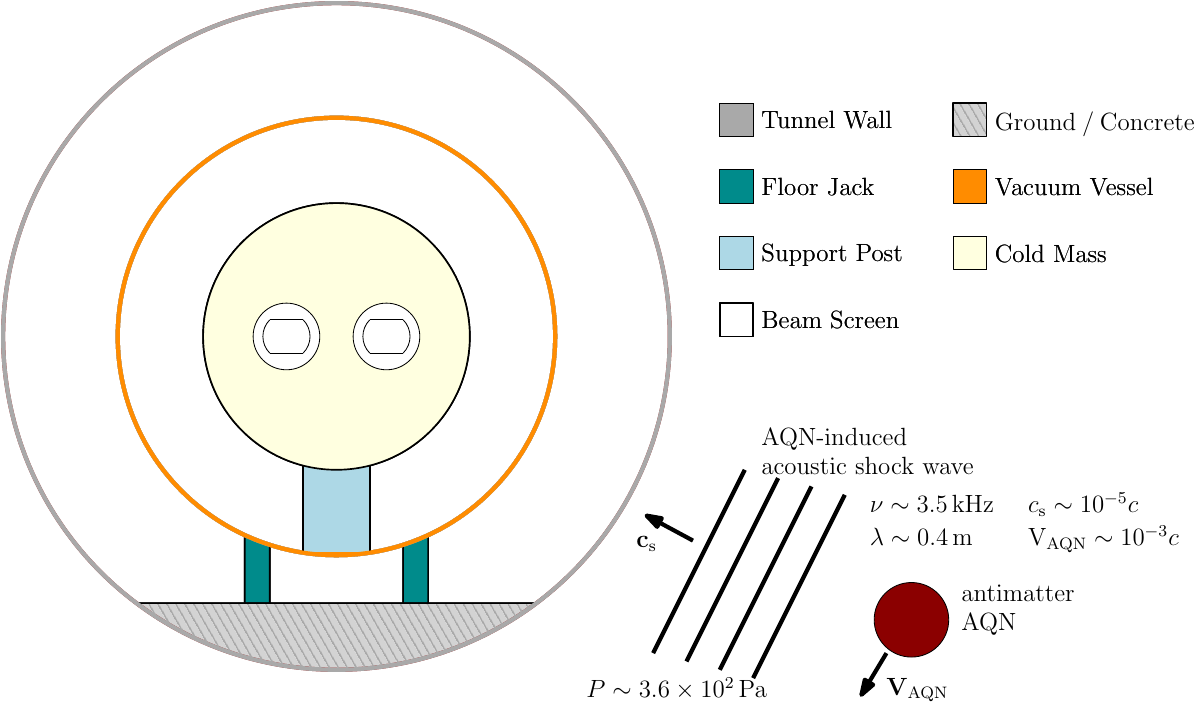}
    \caption{\rev{Simplified model of pressure wave transmission induced by an AQN. Pressure waves propagate through the following layers in order: the concrete tunnel wall, floor jacks, vacuum vessel, support posts, cold mass, and finally the beam screen. Lengths are used to estimate the effect of decoherence. The tunnel wall is treated as a single coherent layer, as it is uniformly driven by a coherent plane wave. The three floor jacks and the three support posts are presumed coherent due to their small size compared to the wavelength of the pressure waves. The configuration is simplified from the original design~\cite{Bourcey:2004un}.
}}
    \label{fig:tunnel}
\end{figure}

Upon transmission, the pressure amplitude is concentrated when the acoustic wave propagates from an outer layer to an inner one with a smaller cross-sectional area:
\begin{equation}
\mathcal{A}_{ij}
=\frac{A_i}{A_j}\,,
\end{equation}
where $A_i$, $A_j$ are the corresponding areas for the outer layer $i$ and the inner  layer $j$, respectively. 

Lastly, we consider the dissipation within a medium $i$. For simplicity, we assume this is dominated by phase decoherence. The initial layer, the tunnel wall, is  uniformly driven by a coherent plane wave. However, the acoustic wave becomes a point source when it transmits from the floor jacks to the vacuum vessel, and from the support posts to the cold mass, and phase decoherence develops. Similar dissipation may happen when the acoustic wave transmits from the cold mass to the two beam screens, which are located off the cylindrical axis. The thickness and circumference of each layer are typically smaller than or comparable to the wavelength of the acoustic wave. 
We estimate the effect of decoherence by the length $L_i$:
\begin{equation}
\mathcal{D}_i(\nu)
=\sqrt{\frac{2c_{{\rm s},i}}{\nu L_i}}\qquad(\nu\gg\nu_{\rm c})\,,
\end{equation}
where $i$ represents the vacuum vessel, cold mass, and the beam screens. We choose $L_i=15{\rm\,m}$ as the characteristic length scale, and the factor of 2 in the formula corresponds to the acoustic wave spreading from the midpoint to its surroundings. The floor jacks and support posts are presumably coherent due to their small size compared 
to the wavelength of the acoustic wave.

Consequently, the pressure transfer function for high-frequency acoustic waves can be estimated as:
\begin{equation}
H_p(\nu)
=\prod_{i=1}^{6}
\mathcal{T}_{i,i+1}\mathcal{A}_{i,i+1}\mathcal{D}_i(\nu)\qquad
(\nu\gg\nu_{\rm c})\,.
\end{equation}
Applying the data in Table~\ref{tab:components}, we obtain:
\begin{equation}
\label{eq:H_p nu approx}
H_p(\nu)
\approx3.18
\left(\frac{D_0}{3.8{\rm\,m}}\right)
\left(\frac{3.5{\rm\,kHz}}{\nu}\right)^{\frac{3}{2}}\qquad
(\nu\gg\nu_{\rm c})
\end{equation}
where $D_0$ is the diameter of the tunnel wall. From this equation, we find that the transmission of high-frequency pressure waves through the LHC tunnel preserves its amplitude very well, and may even gain amplification through the transmission. The most essential factor comes from the effect of area concentration (focusing effect). The tunnel wall acts like a satellite dish and efficiently intercepts the high-frequency waves across its entire surface area and funnels it to a tiny focal point (i.e.\ the beam screens). In contrast, the tunnel wall is too small to ``see'' low-frequency waves. The entire system may wiggle slightly in the field, but it does not collect or focus anything.

While the estimation of the pressure transfer function suggests a potential gain in pressure amplitude, we note that the transmission model in this subsection is highly simplified. For example, we implicitly assume the system has perfect contact and stiff connections between components. In practice, the beam screen may receive less amplification than estimated and only experience pressure comparable to that on the external wall. Even in this scenario, the pressure transmitted is sufficient to trigger a UFO event. Therefore, we will assume $H_p(\nu)\approx1$ in the following discussion, as absorption effects do not play a dominant role for propagating acoustic waves in the kHz range over distances of a few meters.

\begin{table*}[t]
\centering
\caption{\rev{Simplified model of pressure wave transmission induced by an AQN. Pressure waves propagate through the following layers in order: the concrete tunnel wall, floor jacks, vacuum vessel, support posts, cold mass, and finally the beam screen. Lengths are used to estimate the effect of decoherence. The tunnel wall is treated as a single coherent layer, as it is uniformly driven by a coherent plane wave. The three floor jacks and the three support posts are presumably coherent due to their small size compared to the wavelength of the pressure waves. Data sourced from Refs. \cite{Bruning:2004ej,FernandezCano:2006rn,Dupont:1999dma,Benenson:2002qu}.}
}
\label{tab:components}
\begin{tabular}{ccccccc}
\hline\hline
Label & Component & Materials & Density\,$\rm[10^3kg/m^3]$ & Sound speed\,$\rm[km/s]$ & Diameter\,$\rm[m]$ & Length\,[m] \\\hline\hline
0 & Tunnel Wall & Concrete & 2.5 & 3.0 & 3.8 & Coherent \\
1 & Floor Jacks ($\times3$) & Steel & 8.0 & 5.0 & $\lesssim0.2~~~$ & Coherent \\
2 & Vacuum Vessel & Steel & 8.0 & 5.0 & 0.9 & 15\\
3 & Support Posts ($\times3$) & Fiberglass & 2.5 & 3.0 & $\lesssim0.2~~~$ & Coherent \\
4 & Cold Mass & Iron/steel & 8.0 & 5.0 & 0.57 & 15 \\
5 & Beam Screens ($\times 2$) & Steel/copper & 8.0 & 4.5 & 0.04 & 15 \\
\hline\hline
\end{tabular}
\end{table*}

%
}

\section{Event rate}\label{sect:rate}


The estimates presented in the previous section suggest that the passage of an AQN within approximately 100\,km of the LHC may generate acoustic waves. These waves could trigger UFO events, which can be recorded by the BLM system.  The event rate can be estimated using expression (\ref{Phi1}) for the AQN flux  $(\rmd \Phi/ \rmd A)$ on earth, which is expressed as follows:
\begin{equation}
\label{eq:dot N_UFO AQN}
\begin{aligned}
\dot{N}_{\rm UFO}^{\rm(AQN)}
\approx 
A\cdot\left(\frac{\rmd \Phi}{\rmd A}\right)
=4\pi r_{\rm max}^2\left(\frac{\rmd \Phi}{\rmd A}\right)\,,
\end{aligned}
\end{equation}
where $r_{\rm max}\sim100\,$km is the effective range of the AQN-induced acoustic shock wave capable of triggering at least one UFO event.

From the critical condition \eqref{eqs:E_crit etc.}, the activation energy to trigger \textit{one} UFO event would be about $10^6\,$eV. However, one UFO event is not sufficient to distinguish an AQN-induced event from a regular UFO event. The effective event rate should only consider the UFO bursts. Making a conservative estimate, we set the criterion to be $\Delta E_{\rm k}\geq10^7{\rm\,eV}$ that can initiate \textit{multiple} UFO events. From Eq. \eqref{eq:Delta E_k}, this implies:
\begin{equation}
\label{eq:r}
\begin{aligned}
r_{\rm max}
&\approx67{\rm\,km}
\left(\frac{3{\rm\,g\,cm^{-3}}}{\rho_{\rm dust}}\right)
\left(\frac{\langle M_{\rm AQN}\rangle}{100{\rm\,g}}\right)^{\frac{4}{3}}  \\
&\quad \times\left(\frac{L}{10{\rm\,\mu m}}\right)
\rev{
\left(\frac{10^7{\rm\,eV}}{\Delta E_{\rm k,min}}\right)
}\,,
\end{aligned}
\end{equation}
\rev{where $\Delta E_{\rm k,min}$ is the minimum activation energy we assume to trigger multiple UFO events.}
Substituting this value in Eq. \eqref{eq:dot N_UFO AQN}, we obtain:
\begin{equation}
\label{eq:rate}
\begin{aligned}
\dot{N}_{\rm UFO}^{\rm(AQN)}
&\approx0.045{\rm \frac{events}{hour}}
\left(\frac{3{\rm\,g\,cm^{-3}}}{\rho_{\rm dust}}\right)^2
\left(\frac{\langle M_{\rm AQN}\rangle}{100{\rm\,g}}\right)^{\frac{5}{3}}  \\
&\quad\times
\left(\frac{L}{10{\rm\,\mu m}}\right)^2
\left(\frac{\rho_{\rm DM}}{0.3{\rm\,GeV\,cm^{-3}}}\right)  \\
&\quad\times\left(\frac{v_{\rm AQN}}{220{\rm\,km\,s^{-1}}}\right)
\rev{
\left(\frac{10^7}{\Delta E_{\rm k,min}}\right)^2
}
\,.  \\
\end{aligned}
\end{equation}







This estimate highly depends on the sensitivity of the LHC as an acoustic DM detector. It is important to note that this estimation refers to the number of AQN-induced UFO bursts. Consequently, the actual number of UFO events could be several times higher than the rate (\ref{eq:rate}). This is because an acoustic wave may trigger multiple UFO events nearly simultaneously from different locations within LHC tunnel. 

Based on observation \cite{Baer:1379150}, the UFO event rate is less than 5-10 events per hour. Hence, we estimate approximately $(1-10)\%$ of the UFO events may be induced by AQNs. This implies that the AQN-induced UFO events constitute a small subset of the total UFO events. However, as we will discuss below, these events are quite unique and can be easily distinguished from far more frequent conventional UFO events.

\section{Detection perspectives}\label{sect:detection}

In contrast to regular UFO events, the AQN-induced events almost always come with a UFO burst. Such a burst of UFO events happens within a short time (about milliseconds to seconds) and across the entire LHC tunnel (about 8.5\,km in diameter). The burst can be recorded by the BLMs localized at different points along the tunnel ring.

\begin{figure}[h]               
  \centering
\includegraphics[width=1\linewidth]{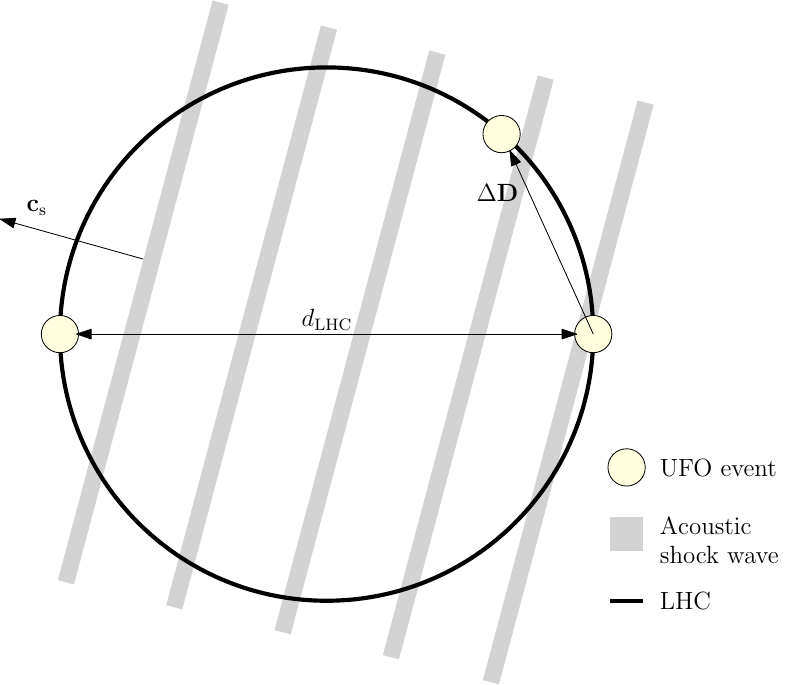}
  \caption{An AQN-induced UFO burst. This acoustic shock wave propagates throughout the entire LHC ring at a velocity $\mathbf{c}_{\rm s}$. It causes mechanical vibration of the dust within the LHC tunnel and triggers several consecutive UFO events. The time interval between two correlated UFO events is determined by the dot product of $\mathbf{c}_{\rm s}$ and the vectorized separation distance $\Delta \mathbf{D}$.}
  \label{fig:burst}
\end{figure}

As shown in Fig. \ref{fig:burst}, the time interval $\Delta t$ between two AQN-induced UFO events can be estimated as:
\begin{equation}
\label{eq:c_s2 c_s cdot Delta D}
\frac{1}{c_{\rm s}^2}|\mathbf{c}_{\rm s}\cdot\Delta\mathbf{D}|
\leq\Delta t\leq\frac{d_{\rm LHC}}{c_{\rm s}}\,, 
\end{equation}
where $\mathbf{c}_{\rm s}$ is the velocity of the acoustic shock wave, $\Delta\mathbf{D}$ is the vector pointing from one UFO event to the other, and $d_{\rm LHC}\approx8.5\,$km is the diameter of the LHC ring. 

Assuming an isotropic distribution of shock wave propagation underground, we obtain the average value of the dot product:
\begin{equation}
\begin{aligned}
\langle |\mathbf{c}_{\rm s}\cdot\Delta\mathbf{D}|\rangle
&\approx\frac{c_{\rm s}\Delta D}{4\pi}
\int_0^{2\pi}d\phi\int_0^{\pi/2} d\theta\,
\sin\theta\cos\theta  \\
&=\frac{1}{4}c_{\rm s}\Delta D\,,
\end{aligned}
\end{equation}
Taking this dot product average and assuming the average distance between two neighbouring BLMs to be $\sim100\,$m, we find the average $\Delta t$ to be
\begin{equation}
\label{eq:6 ms}
6{\rm\,ms}
\left(\frac{\Delta D}{100{\rm\,m}}\right)
\lesssim\Delta t
\lesssim2{\rm\,s}
\left(\frac{d_{\rm LHC}}{8.5{\rm\,km}}\right)\,,
\end{equation}
where we assume the sound speed to be $c_{\rm s}=4{\rm\,km\,s^{-1}}$.
Such correlations, both short-term and long-distance, between multiple UFO events are unique. It can serve as a strong indicator to distinguish AQN-induced UFO events from regular UFO events.

Additionally, nearby seismic stations can be utilized to differentiate these events from regular UFO events. As discussed in  Ref. \cite{Budker:2020mqk}, AQN-induced acoustic shock waves are accompanied by seismic signals, demonstrating a clear correlation between AQN-induced UFO events and local seismic signals. Since 2017, the CERN Seismic Network (deployed in 3 places at CERN: ATLAS, CMS, and Pr{\'e}vessin site) has been in continuous operation in collaboration with the Swiss Seismological Service \cite{Charrondiere:2018ony,Schaumann:2023gst}. This network can effectively distinguish AQN-induced UFO events from other natural or human-related noise.

Furthermore, the Distributed Acoustic Sensing (DAS) and infrasound detectors can also serve as efficient tracers of AQN-induced acoustic waves \cite{Budker:2020mqk}. The DAS instrument is becoming a conventional tool for seismic applications and others. The infrasound is produced by an antimatter AQN propagating through the air: 
\begin{subequations}
\label{eq:P_air r}
\begin{equation}
P_{\rm air}(r)
\sim 0.099{\rm\,Pa}
\left(\frac{100{\rm\,km}}{r}\right)^{\frac{3}{4}} 
\left(\frac{\langle M_{\rm AQN}\rangle}{100{\rm\,g}}\right)^{\frac{2}{3}}\,,
\end{equation}
\begin{equation}
\nu_{\rm air}(r)
\sim 6{\rm\,Hz} 
\left(\frac{100{\rm\,km}}{r}\right)^{\frac{1}{4}}\,.
\end{equation}
\end{subequations}
\rev{
Here we choose the numerical values to match those from the original 
paper~\cite{Budker:2020mqk}, rather than those in Eqs.~\eqref{eqs:P_air etc.}, for the sake of consistency as noted in Sec.~\ref{subsec:Acoustic shock waves emitted by AQNs}.
}

The cross-correlations among different signals and instruments -- including UFO bursts detected by the BLMs, acoustic signals from the seismic network, DAS, and infrasound detectors -- may unambiguously identify AQN-induced UFO events, discriminating 
them from any spurious signals.


\section{Expected sensitivity}
\label{sec:Expected sensitivity}

When observing AQN-induced UFO bursts, it is uncertain whether some signals are genuine or merely regular UFO events that coincidentally occur within a short time frame. We will refer to these events as ``background" events, as they are not related to the AQN-induced UFO events.

For a conservative estimate, the event rate of regular UFOs is no more than 10 events per hour \cite{Baer:1379150}:
\begin{equation}
\label{eq:dot N_UFO reg}
\dot{N}_{\rm UFO}^{\rm(reg)}
\lesssim10{\rm\frac{events}{hour}}
=5.6\times10^{-3}{\rm\frac{events}{2{\rm\,s}}}\,.
\end{equation}

As we will see, the second equality of the equation will be used as an estimate for a ``background" signal   generated by regular UFO events. According to Eq. \eqref{eq:6 ms}, the time interval between two AQN-induced UFO events is no more than 2 seconds. A background  signal must satisfy this criterion. Assuming a Poisson distribution, the probability of observing $n$ or more regular UFO events within 2 seconds is:
\begin{equation}
\textrm{Prob}(n)
=\sum_{k=n}^{\infty}
\frac{\lambda^{k}e^{-\lambda}}{k!}\,,
\end{equation}
where $\lambda=5.6\times10^{-3}$ from Eq. \eqref{eq:dot N_UFO reg}.
Given an measurement time $t_{\rm m}$, the number of $n$- (or more-than-$n$-) correlated events that represent background (noise) is:
\begin{equation}
N_{\rm noi}(n)
={\rm Prob}(n)\times \dot{N}_{\rm UFO}^{\rm(reg)}t_{\rm m}\,.
\end{equation}
Similarly, the number of correlated events that are true AQN-induced signals is:
\begin{equation}
N_{\rm sig}(M_{\rm AQN})
=\eta_{\rm op}\times\dot{N}_{\rm UFO}^{\rm(AQN)}t_{\rm m}\,,
\end{equation}
where $\eta_{\rm op}$ is the fraction of LHC operation time per day. Normally, the operation time can be about 10-14 hours per day. We choose $\eta_{\rm op}\approx50\%$ in this work. The signal-to-noise ratio (SNR) is therefore:
\begin{equation}
\label{eq:SNR}
\textrm{SNR}
=\frac{N_{\rm sig}(M_{\rm AQN})}{\sqrt{N_{\rm noi}(n)}}\,.
\end{equation}
From the dependencies of $N_{\rm sig}$ and $N_{\rm noi}$, the SNR depends on both $\langle M_{\rm AQN}\rangle$ and $n$. The SNR values are presented in Table \ref{tab:SNR} with $t_{\rm m}=360\,$hours, which is equivalent to 30\,days of runs if we assume an LHC operation efficiency of $\eta_{\rm op}=50\%$. Here, we choose $\dot{N}_{\rm UFO}^{\rm(reg)}=10 {\rm\frac{events}{hour}}$. The other basic parameters (including $\xi$, $\eta_{\rm op}$, $\rho_{\rm dust}$, $L$, $\rho_{\rm DM}$, and $v_{\rm AQN}$) following the values used in Eq. \eqref{eq:dot N_UFO AQN}. 

Table \ref{tab:SNR} presents the following observations for a measurement time of $t_{\rm m}=360\,$hours: when two correlated UFO events are recorded, the SNR exceeds 10 for an AQN mass of 50\,g or greater. If three correlated UFO events are detected, the SNR remains above 5 across the entire range of permissible AQN masses. This indicates a remarkably clear signal, even when we adopt a conservative estimate for $\dot{N}_{\rm UFO}^{\rm(reg)}$ as shown in Eq. \eqref{eq:dot N_UFO reg}. As technology continue to advance, the value of $\dot{N}_{\rm UFO}^{\rm(reg)}$ is decreasing, which further enhance the SNR.

\begin{table}[h]
  \centering
  \begin{tabular}{llllll}
    \hline\hline
     & 5\,g & 10\,g & 50\,g & 100\,g & 500\,g \\ \hline
    $n=2$\,: & 0.232 & 0.736 & 10.8 & 34.1  & 499 \\ \hline
    $n=3$\,: & 5.39 & 17.1 &  250 & 794 & $1.16\times10^4$ \\ \hline\hline
  \end{tabular}
  \caption{Signal-to-noise ratio (SNR) as a function of the average AQN mass $\langle M_{\rm AQN}\rangle$ and the number of correlated events $n$ occurring within 2 seconds. The measurement time is chosen to be $t_{\rm m}=360\,$hours. The SNR values are presented in the table based on Eq. \eqref{eq:SNR}. Here we choose the event rate of one UFO event $\dot{N}_{\rm UFO}^{\rm(reg)}=10 {\rm\frac{events}{hour}}$ and LHC operation efficiency $\eta_{\rm op}=50\%$. The other parameters follow the values in Eq. \eqref{eq:rate}. If three correlated UFO events are detected, the resulting signal will be exceptionally clear across the entire range of permitted AQN masses.}
  \label{tab:SNR}
\end{table}

\rev{ 

It is important to emphasize that the SNRs in Table~\ref{tab:SNR} are a qualitative estimation. The quantification of the uncertainties is a complicated analysis as it is related to complexities such as the dust release mechanism, the unknown size and charge distribution of the dust particles, and the actual transfer function of the system. To precisely account for these factors is beyond the scope of this paper, which only serves as the initial proposal to motivate the re-analyses 
of old UFO-LHC records to see if such UFO bursts are present in the data. With this motivation in mind, we have used a very conservative approach in our estimate:
\begin{itemize}
\item While the typical activation energy of a single UFO is about $10^6{\rm\,eV}$ [Eq. \eqref{eq:E_crit}], we assume the critical energy of multiple UFOs to be an order of magnitude larger, $10^7{\rm\,eV}$ [Eq. \eqref{eq:r}];
\item We choose a conservative value of the pressure transfer function 
$H_p(\nu)\approx1$, although it likely provides an enhancement by a factor of 3 [Eq. \eqref{eq:H_p nu approx}];
\item We choose the almost maximal event rate of regular UFOs [10 events per hour, Eq. \eqref{eq:dot N_UFO reg}], which gives the largest background noise to the SNR.
\end{itemize}
We hope this careful approach can justify our estimate of SNR in Table~\ref{tab:SNR}.
}

\section{Conclusion}\label{sect:conclusion}

In this work we argue that 
UFO events can be triggered by an antimatter AQN passing near the LHC within approximately 100\,km. \rev{ In this estimate, we use the basic normalization parameter of the model, the average AQN mass of $\langle M_{\rm AQN}\rangle$. This normalization is based on the assumption that 
the old mysterious and puzzling UV emission in our galaxy is due to DM particles in the form of AQNs, as discussed in Sec.~\ref{basics}.}
 
 The event rate \eqref{eq:rate} of AQN-induced UFOs implies that about $(1-10)\%$ of the UFO events may be induced by AQNs. We propose to search for correlation signals of multiple UFO events across the entire LHC ring that occur within a short time window \eqref{eq:6 ms} (approximately 6\,ms to 2\,s), since this is a unique feature of the AQN-induced events compared to regular UFOs. Assuming a measurement time of about 360 hours, we estimate the $\textrm{SNR}>10$ for $\langle M_{\rm AQN}\rangle\geq50{\rm\,g}$ if two correlated UFO events are recorded; and $\textrm{SNR}>5$ for $\langle M_{\rm AQN}\rangle\geq5{\rm\,g}$ if three correlated UFO events are recorded. This implies that the LHC can serve as a practical, broadband Large Acoustic Detector for AQNs.

This proposal is significantly more efficient \rev{ for the AQN DM than traditional acoustic detector networks as originally proposed in Ref.~\cite{Budker:2020mqk}.
Existing seismic networks are not optimized for the kilohertz frequency range. Detecting kHz acoustic waves would require specialized infrastructure, such as DAS detectors, as discussed in Sec.~\ref{sect:detection}. However, to mitigate anthropogenic noise, even a modest array of tens of sensors deployed 100 meters underground across a wide area would require a multi-million dollar investment. In contrast, the LHC serves as a high-SNR detector with virtually no additional infrastructure costs. It is a pre-existing 27\,km continuous ring built 100\,m underground with approximately 4000 BLMs. It benefits from a low noise floor at high frequencies thanks to the acoustic shielding provided by the vacuum vessel. Rather than requiring the continuous operation of an expensive dedicated network, the primary requirement for detection with the LHC is data mining.
}

Generally, our proposal in this work is not limited to the LHC but is potentially applicable to all particle accelerators, \rev{potentially forming a global acoustic detector network.} However, smaller accelerators tend to have a lower UFO trigger probability, which implies a higher critical energy $\Delta E_{\rm k}$ to initiate correlated events. The performance of smaller accelerators as broadband acoustic detectors depends on the local dust particle size distribution. A detailed scaling analysis is beyond the scope of this study but would be a useful direction for future work.

\rev{
An additional improvement is to utilize infrasound networks for cross-validation, as discussed in Sec.~\ref{sect:detection}. As presented in Eq.~\eqref{eq:P_air r}, an AQN emits low-frequency (6\,Hz) infrasound waves during its propagation in the air. A notable example is the Elginfield Infrasound Array (ELFO), as  mentioned in Sec.~\ref{sect:transfer} and discussed in more detail in Appendix~\ref{sect:skyquakes}.
}

%
%
%
%

The AQN DM model presented here is consistent with current cosmological, astrophysical, satellite, and ground-based observations. Indeed, it may help illuminate several long standing puzzles discussed above. If validated by the proposed LHC searches, through selection of the distinctive UFO-burst signature and its correlation with seismic and infrasound signals, this framework could fundamentally change our understanding of DM and its role in the cosmos. Within the AQN picture, UFO-burst events would constitute direct terrestrial manifestations of DM and could enable \textit{direct} studies of DM at the LHC, complementing the \textit{indirect} AQN-induced signatures described earlier.

\section*{Acknowledgements}
This research was supported in part by the Natural Sciences and Engineering Research Council of Canada. 
We thank Philippe Belanger, Rudiger Schmidt, Christoph Wiesner, and Daniel Wollmann for their discussions, clarifications, and explanations regarding the observed UFO events.
We are also thankful to Surya Sundar Raman for the helpful discussions during the initial stages of the project.

\appendix

\section{Skyquakes as AQN-Induced Events}\label{sect:skyquakes}





\rev{ The main goal of this Appendix is to briefly review the nature of the shock waves which are the source of rare LHC UFO events as proposed in this work. The basic formulae relevant for this work have already been reviewed in Sec.~\ref{subsec:Acoustic shock waves emitted by AQNs}. In this Appendix we would like to give a historical overview of the nature of these rare acoustic natural phenomena.}
    
For many years, numerous mysterious and puzzling observations have been collected that cannot be explained by conventional physics despite extensive efforts. In this discussion, we focus on the phenomenon known as skyquakes, which have been reported for centuries without any viable explanations from the standpoint of conventional physics \cite{skyquakes}. Skyquakes are extremely rare acoustic events that sound like a cannon shot or a sonic boom coming from the sky. A description provided by a meteorologist in a TV interview illustrates this phenomenon well \cite{skyquakes-meteorologist}. The main message from this short interview is that skyquakes cannot be attributed to seismic activities or meteor-related events, which are routinely recorded around the world. These events cannot be linked to any human activities that are also documented. It is noteworthy that such events have been observed for centuries across different countries, each with varying environmental features. These events cannot be explained by military aircraft, as reports of skyquakes date back long before the advent of supersonic flights. 
 
We also want to highlight a powerful and rare event that occurred in Alabama on a clear day in 2017, when a sound was heard across 15 counties \cite{NASA}. There were no meteor-activity reports associated with this event; no visible lights, which typically accompany meteoroids, were recorded. However, there were reports of vapor trails. It is important to note that the presence of vapor is expected and even predicted by the proposal \cite{Budker:2020mqk} that associates skyquakes with AQN-induced acoustic shock waves. This is because most of the energy released, as described by Eq. (\ref{T}), is in the form of X-rays, which will heat the surrounding material.  Therefore, the absence of reported meteor activity (no visible light recorded) in this event, alongside the presence of vapor trails, is not surprising \cite{NASA}.  

We want to clarify that the information about skyquakes is not primarily disseminated through gossip, such as the aforementioned TV interviews \cite{skyquakes-meteorologist, NASA}, but rather through solid scientific studies. 
An example of scientific analysis is the powerful and rare event that took place on July 31, 2008, which was properly recorded by the ELFO near London, Ontario, Canada \cite{ELFO}. 
The detection of infrasound was accompanied by nonobservation of any meteors by an all-sky camera network, effectively ruling out a conventional meteor source. Additionally, no meteorites were discovered in the active area. Besides infrasound, seismic impulses were also detected shortly thereafter as ground-coupled acoustic waves around Southwestern Ontario and Northern Michigan. This event was treated as an AQN-induced event by Ref. \cite{Budker:2020mqk}, and it was argued that the energetic properties, infrasound frequency characteristics, and other attributes of the event are consistent with the observed ELFO phenomenon. It was estimated in Ref. \cite{Budker:2020mqk} that the baryon charge of the ELFO event is around 
$M_{\rm AQN}\approx2\,$kg, which is significantly higher than a typical AQN size.  According to the findings represented in Eqs. (\ref{Phi1}) and (\ref{eq:f(B)}), ELFO-like powerful events are indeed very rare. 
  
As argued in Ref. \cite{Budker:2020mqk}, infrasound waves are generated by the propagation of fast moving AQNs through the atmosphere. However, there will be minimal electromagnetic radiation in the visible frequency range; most emissions will manifest as hard X-rays due to the extremely high internal temperature (\ref{T}) of the AQN. Upon reaching the Earth's surface at an enormous speed of approximately $\sim 10^{-3}c$, the AQN will continue to propagate into the deeper regions of the Earth. This journey will be accompanied by annihilation events, which will heat the surrounding material along its path. 
As the AQN travels, it generates seismic waves. The events induced by the AQN can be distinguished from background noise because there will be a correlation between the infrasound acoustic waves and the seismic waves. This correlation forms the basis for investigating these rare AQN-induced events, allowing researchers to discriminate them from noise and other spurious signals~\cite{Budker:2020mqk}. \rev{ The key element of 
the present work is that the entire ring of the LHC may serve as a unique DM detector for such rare natural events as argued in Sec.~\ref{sect:detection}.}


\rev{

\section{Related terrestrial observations}
\label{app:Related terrestrial observations}

There are terrestrial observations that may support the DM in the form of strongly interacting AQNs, rather than conventional WIMPs or axions. Below, we list recent studies of puzzling Earth-based phenomena that might result from AQN annihilation with surrounding matter. We focus on representative events recorded by modern instruments.

Similar to the AQN-induced UFO-LHC  bursts, the Telescope Array (TA) reported mysterious bursts in which at least three air showers were recorded within 1\,ms. Such bursts cannot be explained by conventional high energy cosmic rays (CRs), since the rate at which two or more consecutive energetic CRs would hit the same area is expected to be on the order of days or even months \cite{Abbasi:2017rvx,Okuda_2019}. These TA bursts were found to be associated with lightning activity in the area. In our previous works \cite{Zhitnitsky:2020shd,Liang:2021wjx}, we proposed that these puzzling events could be related to AQN-induced phenomena. Notably, we used the same formula \eqref{Phi1} to estimate the rate of the TA bursts as we used to calculate the event rate of AQN-induced UFOs [Eq. \eqref{eq:rate}]. The rarity of these TA burst (only a few events in ten years of observation) can be explained by the low expected rate of AQN events.

Likewise, ``Exotic Events" recorded by the AUGER collaboration \cite{PierreAuger:2021int,2019EPJWC.19703003C,Colalillo:2017uC} are also associated with thunderstorm activity in the area. These events cannot be explained by canonical CR modelling. However, these Exotic Events recorded by the AUGER collaboration can be accounted for within the AQN framework, as argued in \cite{Zhitnitsky:2022swb}. Furthermore, the rarity of these Exotic Events is explained in the AQN framework by the same formula \eqref{Phi1}, and this prediction is consistent with the event counts reported by the Auger Collaboration.

Rare and unusual CRs were also observed in Antarctica by \textsc{ANITA}. In particular, two anomalous events with noninverted polarity were recorded, which correspond to upward-going CRs at relatively steep arrival angles \cite{Gorham:2016zah,Gorham:2018ydl}. Although the two events can be interpreted as $\nu_\tau$ neutrinos with energies of order EeV, such an interpretation is exceedingly unlikely according to standard model physics. As proposed in Ref. \cite{Liang:2021rnv}, these two anomalous events can be naturally explained within the AQN framework.

Lastly, we choose a recent study \cite{Zhitnitsky:2024jnk} that examines anomalies of the Earth's atmosphere. These anomalies include unexpected correlations between temperature variations in the stratosphere and the total electron content of the atmosphere, among other mysterious correlations \cite{Zioutas_2020,Argiriou:2025xsq}. This suggests that the entire globe can be thought of as a single large detector. Decades of data collection provide statistically significant evidence for these observed correlations.  However, no conventional explanations have been proposed for these mysterious correlations.  The study \cite{Zhitnitsky:2024jnk} argues that the observed correlations may be the result of processes induced by AQNs. 

}

\rev{

\section{Pressure transmission coefficient}
\label{app:Pressure transmission coefficient}

Here we derive the pressure transmission formula in Eq.~\eqref{eq:T_ij}. Readers may also refer to textbooks such as Refs.~\cite{landau-fluid,Benenson:2002qu}.

We start with Newton's second law:
\begin{equation}
\label{eq:rho partial u partial t}
\rho\frac{\partial \mathbf{u}}{\partial t}
=-\boldsymbol{\nabla} p
\end{equation}
where $\mathbf{u}$ is the vibration velocity of an atom (or a molecule) in the medium. For sinusoidal plane waves, one can Fourier transform Eq.~\eqref{eq:rho partial u partial t} into
\begin{equation}
-i\mathbf{k}p=-i\omega\rho\mathbf{u}\,,
\end{equation}
where $k$ and $\omega$ are the wave number and angular frequency of the acoustic wave. 
This implies that $\mathbf{k}$ and $\mathbf{u}$ point along the same direction, namely 
$\mathbf{\hat{k}}=\mathbf{\hat{u}}$. This implies
\begin{equation}
\label{eq:mathbf u}
\mathbf{u}
=\frac{p}{Z}\mathbf{\hat{u}}
\end{equation}
where the characteristic sound impedance is defined to be $Z=\rho c_{\rm s}$ and $c_{\rm s}=\frac{\omega}{k}$ is the sound speed.

For two media in perfect contact, the jump in the normal component ($\mathbf{\hat{n}}$) of the velocity across the interface is zero:
\begin{equation}
\label{eq:mathbf u_i mathbf u_r cdot mathbf hat n}
(\textbf{u}_{\rm i}+\mathbf{u}_{\rm r})\cdot\mathbf{\hat{n}}
=\mathbf{u}_{\rm t}\cdot\mathbf{\hat{n}}\,,
\end{equation}
where the subscripts i, r, t correspond to the incident, reflected, and transmitted components of the acoustic wave. For normal incidence, 
Eqs.~\eqref{eq:mathbf u} and \eqref{eq:mathbf u_i mathbf u_r cdot mathbf hat n} can be combined and simplified to:
\begin{equation}
\label{eq:p_i Z_1 p_r Z_1}
\frac{p_{\rm i}}{Z_1}-\frac{p_{\rm r}}{Z_1}
=\frac{p_{\rm t}}{Z_2}\,,
\end{equation}
where we assume the acoustic wave goes from medium 1 to medium 2, and $Z_1$ and $Z_2$ are the corresponding characteristic sound impedances. Also, the continuity of pressure implies:
\begin{equation}
\label{eq:p_i p_r}
p_{\rm i}+p_{\rm r}
=p_{\rm t}\,.
\end{equation}
Solving both equations~\eqref{eq:p_i Z_1 p_r Z_1} and \eqref{eq:p_i p_r} gives the reflection ($\mathcal{R}_{12}$) and transmission ($\mathcal{T}_{12}$) coefficients of the pressure from medium 1 to medium 2:
\begin{subequations}
\begin{equation}
\mathcal{R}_{12}
=\frac{p_{\rm r}}{p_{\rm i}}
=\frac{Z_2-Z_1}{Z_1+Z_2}\,,
\end{equation}
\begin{equation}
\mathcal{T}_{12}
=\frac{p_{\rm t}}{p_{\rm i}}
=\frac{2Z_2}{Z_1+Z_2}\,.
\end{equation}
\end{subequations}
This reproduces the pressure transmission coefficient~\eqref{eq:T_ij} from the main body of the text.

}

\bibliography{UFO-LHC}

\end{document}